\documentclass[reqno]{nrc2}
\usepackage{graphicx}
\usepackage{amsmath}
\usepackage{amssymb}
\usepackage{bm}
\usepackage{times}
\usepackage{mathptmx}
\usepackage{accents}
\usepackage{color}

\DeclareMathOperator{\erfc}{erfc}

\begin{document}

\title{Comment on ``The velocity field due to an oscillating plate in an Oldroyd-B fluid'' by C.C. Hopkins and J.R. de Bruyn [Can. J. Phys. 92, 533 (2014)]}

\author[I.C. Christov]{Ivan C. Christov}
\address{Theoretical Division and Center for Nonlinear Studies, Los Alamos National Laboratory, Los Alamos, NM 87545, USA}
\correspond{christov@alum.mit.edu}

\shortauthor{I.C. Christov}

\begin{abstract}
We correct certain errors and ambiguities in the recent pedagogical article by Hopkins and de Bruyn. The early-time asymptotics of the solution to the transient version of Stokes' second problem for an Oldroyd-B fluid in a half-space is presented, as an Appendix, to complement the late-time asymptotics given by Hopkins and de Bruyn.
\end{abstract}

\begin{resume}
En Fran\c{c}ais...
\end{resume}

\maketitle

In a recent pedagogical article, Hopkins and de Bruyn \cite{hdb14} present a treatment of the unsteady boundary layer created by the oscillatory motion of a plate in a certain viscoelastic fluid. Unfortunately, several errors and ambiguities appear in their discussion. Given that the goal of \cite{hdb14} is to be an accessible and, more importantly, useful introduction to the topic for undergraduate students, who are unfamiliar with the subtleties of non-Newtonian fluid mechanics, it is crucial to correct the errors and clarify the ambiguities in \cite{hdb14}. 

\begin{enumerate}
\item In the Introduction and in Section 3 of \cite{hdb14}, the authors say that they are solving for a ``steady-state velocity field.'' This statement is ambiguous since the final velocity field quite clearly depends on time, hence it is unsteady. It is more accurate to say that the solution sought and found is the {late-time} (\emph{post-transient}) velocity field.

\item The authors mention, in passing, above Eq.\ (1) in \cite{hdb14} that $\bm{\tau}$ represents the shear (\emph{deviatoric}) part of stress tensor $\bm{\sigma}$. This is important to emphasize since non-Newtonian constitutive relations are, by convention, written in terms of $\bm{\tau}$, while it is $\bm{\sigma}$ that enters the conservation of linear momentum equation. The difference between $\bm{\sigma}$ and $\bm{\tau}$ is the isotropic stress (constitutively indeterminate for incompressible fluids), usually identified as the pressure $p$, i.e., $\bm{\sigma} = -p\bm{I} + \bm{\tau}$, where $\bm{I}$ is the identity tensor. Note that this is not just a pedantic remark since the general form of the right-hand side of Eq.~(4) in \cite{hdb14} is $\nabla\cdot\bm{\sigma}$, which then makes clear to the reader the origin of term $-\nabla p$.

\item Equation (3) in \cite{hdb14} is missing a term. Since, at this point in the discussion, the authors have not restricted to incompressible fluids or isochoric flows, then the correct form of the upper-convected invariant time derivative of a \emph{tensor} $\bm{A}$ (see Sec.~3(a) in \cite{o50} or Eqs.~(8.32.2) and (8.32.3) in \cite{a62}) is
\begin{equation}
\accentset{\bigtriangledown}{\bm{A}} = \frac{\partial \bm{A}}{\partial t} + (\bm{u}\cdot\nabla)\bm{A} - (\nabla\bm{u})^\top\bm{A} - \bm{A}\nabla\bm{u} + (\nabla\cdot\bm{u})\bm{A},
\end{equation}
where the last term in the equation above is missing from Eq.\ (3) in \cite{hdb14}.

\item Section 2 in \cite{hdb14} concludes with the following incorrect statement:
``As an aside, we note that the Cattaneo equation for heat transport has exactly the same form as (2) [21]. In this case, there are no physical restrictions on the values of $\lambda_1$ and $\lambda_2$.'' In the latter, [21] is \cite{jp89} below. Referring to Eq.~(1.2) (``Cattaneo's equation'') in \cite{jp89}, it should be clear that the \emph{formal} analogy between the constitutive equation relating the {deviatoric} stress $\bm{\tau}$ to the rate of strain $\dot{\bm{\gamma}}$ and the constitutive equation relating the heat flux $\bm{q}$ to the temperature gradient $\nabla T$ only holds for the case when $\lambda_2=0$ (using the notation of Eq.\ (2) in \cite{hdb14}). It is also worth noting that the model attributed to Cattaneo \cite{c48} was suggested earlier by Maxwell \cite{m67}. Finally, the value of $\lambda_1$ is not unrestricted in the heat conduction context. It is known that if $\lambda_1$ is ``too large,'' then the Maxwell--Cattaneo heat conduction model can lead to the violation of the second law of thermodynamics \cite{r92}.

\item Throughout the article, Eq.~(4) in \cite{hdb14} is referred to as ``the Navier--Stokes equation'' but this only true for the Newtonian constitutive relation. Equation (4) in \cite{hdb14} is a much more general relation known as \emph{Cauchy's First Law of Motion}, expressing the conservation of linear momentum for a continuum (see, e.g., Section III.6 in \cite{t91}). Meanwhile, ``the Navier--Stokes equation'' usually refers to the equation resulting from substituting   the Newtonian constitutive relation (the so-called ``linearly viscous fluid'') into Cauchy's equation (see, e.g., Eq.~(IV.4-25) in \cite{t91} and the discussion thereafter).

\item Below Eq.~(12) in \cite{hdb14}, the authors emphasize that the ``constitutive relation is nonlinear''. While, generally speaking, this statement has some merit, nonlinear terms such as $\bm{u}\cdot\nabla\bm{u}$ vanish \emph{identically} for the assumed unidirectional velocity field given in Eq.~(11) of \cite{hdb14}. Moreover, the extra assumptions regarding the amplitude of the oscillations and the (non-)excitation of higher-harmonics are neither used nor needed to obtain the \emph{linear} ordinary differential equation (Eq.~(32) in \cite{hdb14}) governing the spatial structure of the post-transient velocity field. Hence, we find these several sentences below Eq.~(12) in \cite{hdb14} to be unclear and potentially confusing for the uninitiated reader.

\item Above Eq.~(22) in \cite{hdb14}, ``the constitutive relation, (3)'' should be replaced by ``the constitutive relation, (15)''. Furthermore, in the top-left entry of the matrix on the left-hand side of Eq.~(22) in \cite{hdb14}, ``$\tau_{yx}-\tau_{xy}$'' should be replaced by ``$\tau_{yx}+\tau_{xy}$''. Similarly, ``$\tau_{yx}-\tau_{xy}$'' should be replaced by ``$\tau_{yx}+\tau_{xy}$'' in the third term on the left-hand side of Eq.\ (23) in \cite{hdb14}.

\item In Figs.~2 and 3 in \cite{hdb14}, (dimensionless) values of $\lambda_1$ and $\lambda_2$ up to 10 are used. Although this is fine for illustrating the qualitative nature of the curves plotted, it is important to note that the two relaxation times are expected to be small (especially compared to the inverse plate frequency $1/\omega$ used as the time scale in \cite{hdb14}), based on available experimental data \cite{ts53}.

\item Last, but not least, a number of references cited in \cite{hdb14} are \emph{mathematically erroneous}, hence we \emph{strongly} advise readers to avoid these papers. It behooves the community to cease the promulgation of these incorrect results that plague the modern fluid mechanics literature. Essentially, the mistake can be boiled down to a sloppy enforcement of the start-up condition for the transient, unsteady problem. For example, in \cite{cj12}, it was shown that Refs.\ [7,8,14] from \cite{hdb14} contain mathematical errors; in \cite{c13}, it was shown that Ref.\ [9] from \cite{hdb14} contains mathematical errors. Meanwhile, Ref.\ [15] in \cite{hdb14} makes use, \emph{without attribution}, of the mathematical technique introduced in \cite{cc10}. Although Refs.\ [11,13] from \cite{hdb14} do not immediately fall within the classes of mathematical errors described in \cite{cj12,c13,cc10,cj09}, we caution the reader against assuming that the results therein are correct.

\end{enumerate}

\section*{Acknowledgement}
I.C.C.\ was supported, in part, by the LANL/LDRD Program through a Feynman Distinguished Fellowship. Los Alamos National Laboratory (LANL) is operated by Los Alamos National Security, L.L.C.\ for the National Nuclear Security Administration of the U.S.\ Department of Energy under Contract No.\ DE-AC52-06NA25396. The author thanks Dr.\ P.M.\ Jordan for suggesting the calculation in the Appendix.

\numberby{equation}{section}
\setcounter{equation}{0}
\appendix*
\section{Early-Time Asymptotics}

In this Appendix, we seek a solution to Eqs.\ (14)--(16) in \cite{hdb14}, without making the assumption that the velocity is separable, namely we do \emph{not} assume $u(y,t) = f(y)\mathrm{e}^{\mathrm{i}t}$ as in Eq.\ (12) in \cite{hdb14}. The exact solution to the transient version of Stokes' second problem for an Oldroyd-B fluid in a half-space is readily established from the solution to the transient version of Stokes' first problem for an Oldroyd-B fluid in a half-space, which is given by Eq.~(7) in \cite{cj09} (keeping in mind that $\alpha$ and $\alpha_t$ therein correspond to $\lambda_1$ and $\lambda_2$ herein) using the convolution theorem for the Laplace transform. The resulting expressions are lengthy, hence, for the sake of brevity, we do not provide them here. (It is important to note, however, that the solution to the transient version of Stokes' second problem for an Oldroyd-B fluid in a half-space given by Eq.~(3.10) in \cite{afm06} is \emph{erroneous} for the reasons listed in \cite{cj12}. Hence,  carrying out the tedious integration and simplification may have scientific value.)

On the other hand, noting that small-$t$ asymptotics correspond to large-$p$ asymptotics in the Laplace transform domain (see, e.g., Sec.~2.2 in \cite{bsj08}), we can obtain the early time behavior of the transient version of Stokes' second problem for an Oldroyd-B fluid in a half-space. To this end, specializing to the $\beta=0$ case of Eq.~(1a) in \cite{cj09}, taking $u(0,t) = H(t)\cos t$ as the first boundary condition in Eq.~(1b) of \cite{cj09} and applying the Laplace transform in $t$, we obtain the subsidiary ordinary differential equation
\begin{subequations}\begin{align}
&\left(\frac{1+\lambda_1 p}{1+\lambda_2 p}\right)p\bar{u} = \frac{\partial^2 \bar{u}}{\partial y^2},\qquad y\in(0,\infty),\\ 
&\bar{u}(0,p) = \frac{p}{p^2 + 1},\qquad \bar{u}(\infty,p) = 0.
\end{align}\label{eq:subsidiary}\end{subequations}
where $\bar{u}$ denotes the (temporal) Laplace transform of $u$ with parameter $p$. Solving Eqs.~\eqref{eq:subsidiary} is straighforward:
\begin{equation}
\bar{u}(y,p) = \frac{p}{p^2+1} \exp\left\{-y\sqrt{\frac{(1+\lambda_1 p)p}{1+\lambda_2 p}}\right\}.
\end{equation}
Expanding for $p\to\infty$, we find
\begin{multline}
\bar{u}(y,p) = \left[\frac{1}{p} + \mathcal{O}\left(\frac{1}{p^3}\right) \right]\\
\times \exp\left\{-y\chi \sqrt{p} + \frac{y \chi \Lambda}{2\sqrt{p}} + \mathcal{O}\left(\frac{1}{p^{3/2}}\right) \right\}\quad (p\to\infty),
\end{multline}
where $\chi = \sqrt{\lambda_1/\lambda_2}$ and $\Lambda = (\lambda_1-\lambda_2)/(\lambda_1\lambda_2)$, whence
\begin{multline}
\bar{u}(y,p) = \left[\frac{1}{p} + \mathcal{O}\left(\frac{1}{p^3}\right) \right]\\
\times \exp\left\{-y\chi \sqrt{p}\right\} \left[1 + \frac{y \chi \Lambda}{2\sqrt{p}} + \mathcal{O}\left(\frac{1}{p}\right)\right]\quad (p\to\infty),
\end{multline}
Inverting the Laplace transform term-by-term, we obtain
\begin{multline}
u(y,t) \simeq \erfc\left(\frac{y\chi}{2\sqrt{t}}\right)\\ + \frac{y\chi\Lambda}{2} \left[2\sqrt{\frac{t}{\pi}}\exp\left(-\frac{y^2\chi^2}{4t}\right) - y\chi\erfc\left(\frac{y\chi}{2\sqrt{t}}\right)\right]
\end{multline}
as the early time (small-$t$) asymptotics. Clearly, the initial development of the boundary layer looks diffusive with a correction proportional to $\Lambda$ due to the non-Newtonian nature of the Oldroyd-B fluid.

\BalanceColumns[0.5]

\end{document}